\documentclass[aps,twocolumn,prb,showpacs,superscriptaddress]{revtex4}
\usepackage{epsfig}
\usepackage{bm}
\def\be{\begin{equation}}
\def\ee{\end{equation}}
\def\bea{\begin{eqnarray}}
\def\eea{\end{eqnarray}}

\begin{document}
\title{Ferrimagnetism of the Heisenberg Models 
on the Quasi-One-Dimensional 
Kagome Strip Lattices}
\author{Tokuro Shimokawa}
\affiliation{Graduate School of Material Science, University of Hyogo, Kamigori, Hyogo 678-1297, Japan}
\author{ Hiroki Nakano}
\affiliation{Graduate School of Material Science, University of Hyogo, Kamigori, Hyogo 678-1297, Japan}

\email{E-mail address:  t.shimokaw@gmail.com } 
\email{E-mail address: hnakano@sci.u-hyogo.ac.jp}

\date{\today}

\begin{abstract}
We study 
the ground-state properties of the $S=1/2$ Heisenberg models 
on the quasi-one-dimensional kagome strip lattices
by the exact diagonalization and 
density matrix renormalization group methods. 
The models with two different strip widths 
share the same lattice structure in their inner part 
with the spatially anisotropic two-dimensional kagome lattice. 
When there is no magnetic frustration, 
the well-known Lieb-Mattis ferrimagnetic state is realized 
in both models. 
When the strength of magnetic frustration is increased, 
on the other hand, the Lieb-Mattis-type ferrimagnetism 
is collapsed. 
We find that there exists 
a non-Lieb-Mattis ferrimagnetic state 
between the Lieb-Mattis ferrimagnetic state and 
the nonmagnetic ground state. 
The local magnetization clearly shows 
an incommensurate modulation with long-distance periodicity 
in the non-Lieb-Mattis ferrimagnetic state. 
The intermediate non-Lieb-Mattis 
ferrimagnetic state occurs irrespective of strip width, 
which suggests that the intermediate phase 
of the two-dimensional kagome lattice is also 
the non-Lieb-Mattis-type ferrimagnetism.
\end{abstract}


\maketitle
\section{Introduction}
Ferrimagnetism is a fundamental phenomenon 
in the field of magnetism. 
The simplest case of the conventional ferrimagnetism 
is the ground state of the mixed spin chain.
For example, there is an ($s$, $S$)=(1/2, 1)  mixed spin chain 
with a nearest-neighbor antiferromagnetic (AF) isotropic 
interaction. 
In this system, 
the so-called Lieb-Mattis (LM)-type ferrimagnetism\cite{Lieb, 
Marshall, Takano, Okamoto, Tonegawa, Sakai} is realized 
in the ground state 
because two different spins are arranged alternately 
in a line owing to the AF interaction. 
Since this type of ferrimagnetism has been studied extensively, 
the magnetic properties 
and the occurrence mechanism of the LM ferrimagnetism 
are well known.
In particular, 
the ferrimagnetism in the quantum Heisenberg spin model 
on the bipartite lattice without frustration 
is well understood within 
the Marshall-Lieb-Mattis (MLM) theorem\cite{Lieb, Marshall}. 
In the LM ferrimagnetic ground state, 
the spontaneous magnetization occurs and the magnitude 
is fixed to a simple fraction of the saturated magnetization. 

In recent years, a new type of ferrimagnetism, 
which is clearly different from LM ferrimagnetism, 
has been found in the ground state 
of several one-dimensional frustrated Heisenberg spin 
systems\cite{PF1, PF2, PF3, PF4, PF5, Shimokawa, Shimokawa2}. 
The spontaneous magnetization in this new type of ferrimagnetism 
changes gradually with respect to the strength of frustration. 
The incommensurate modulation with long-distance periodicity 
in local magnetizations 
is also a characteristic behavior of the new type 
of ferrimagnetism.
Hereafter, we call the new type of ferrimagnetism 
the non-Lieb-Mattis (NLM) type. 
The mechanism of the occurrence of the NLM ferrimagnetism 
has not yet been clarified. 

On the other hand, some candidates of the NLM ferrimagnetism among the two-dimensional (2D) systems, 
for example, the mixed-spin $J_{1}$-$J_{2}$ Heisenberg model on the square lattice\cite{2D-candidate1} and the $S=1/2$ Heisenberg model on the Union Jack lattice\cite{2D-candidate2}, were reported. 
These 2D frustrated systems have the intermediate ground state, namely, the ``canted state'', in which the spontaneous magnetization is changed when the inner interaction of the system is varied.
It has not been, however, determined whether the incommensurate modulation with long-distance periodicity exists in the local magnetization of the canted state owing to the difficulty of treating these 2D frustrated systems numerically and theoretically. Therefore, the relationships between the canted states of these 2D frustrated systems and the NLM ferrimagnetic state 
are still unclear.

\begin{figure*}[t]
\begin{center}
\includegraphics[width=15cm]{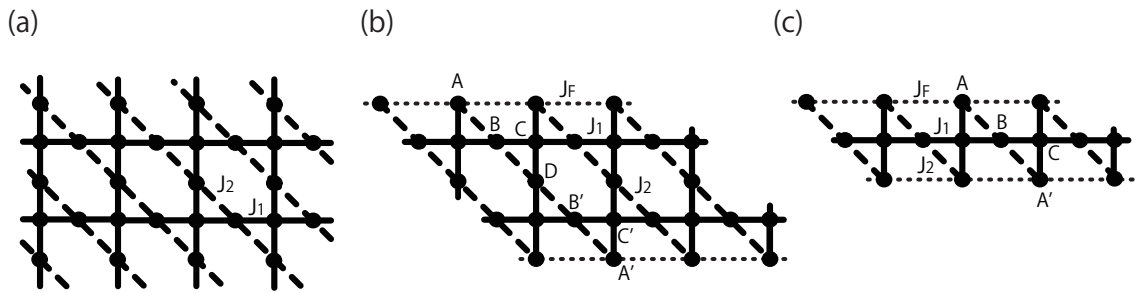}
\caption{
Structures of the lattices: 
the spatially anisotropic kagome lattice (a) and 
the quasi-one-dimensional kagome strip lattices (b) and (c) 
with different widths. 
An $S=1/2$ spin is located at each site denoted 
by a black circle.
Antiferromagnetic bonds $J_{1}$ (bold straight line) 
and $J_{2}$ (dashed line), 
and ferromagnetic bond $J_{\rm F}$ (dotted line). 
The sublattices in a unit cell of lattice (b) are represented 
by A, ${\rm A}^{\prime}$, B, ${\rm B}^{\prime}$, 
C, ${\rm C}^{\prime}$, and D.
The sublattices in a unit cell of lattice (c) are represented 
by A, ${\rm A}^{\prime}$, B, and C. 
}
\label{fig1}
\end{center}
\end{figure*}

Under these circumstances, recently, another candidate of the NLM ferrimagnetism among the 2D systems was reported in ref.~\ref{kagome-2D} 
in which the $S=1/2$ Heisenberg model on the spatially anisotropic kagome lattice depicted in Fig. \ref{fig1}(a) was studied. 
A region of the intermediate-magnetization states is observed between the LM ferrimagnetism that is rigorously proved by the MLM theorem\cite{Lieb, Marshall} 
and the nonmagnetic state of the spatially isotropic kagome-lattice antiferromagnet in the absence of magnetic field\cite{Lecheminant, Waldtmann,Hida_kagome,Cabra,Honecker0,Honecker1,Cepas, Spin_gap_Sindzingre,HN_Sakai2010,Sakai_HN_PRBR,HN_Sakai2011}. 
The local magnetization in the intermediate state of the kagome lattice was investigated by the exact diagonalization method, and it was reported that the local magnetization greatly depends on the position of the sites, although it is difficult to determine clearly whether the incommensurate modulation with long-distance periodicity is present.
This result leads to the high expectation that the intermediate state of the spatially anisotropic kagome lattice is the NLM ferrimagnetic state.
Additional research is desirable 
to conclude that the intermediate state of this 2D system 
is the NLM ferrimagnetism.

In this paper, we study the ground-state properties 
of the $S=1/2$ Heisenberg models 
on the quasi-one-dimensional (Q1D) kagome strip 
lattices depicted in Figs.~\ref{fig1}(b) and \ref{fig1}(c) 
instead of the 2D lattice depicted in Fig.~\ref{fig1}(a). 
Note that the inner parts of the lattices 
in Figs.~\ref{fig1}(b) and \ref{fig1}(c) 
are common to a part of the 2D lattice in Fig.~\ref{fig1}(a). 
We also note that 
the lattice shapes of strips in the present study 
are different from some kagome strips (chains) 
studied in refs. 
\ref{kagome-strip1}-\ref{kagome-strip3}, 
where the nontrivial properties of kagome antiferromagnets 
were reported.

According to the study in ref. \ref{kgm-stripe}, 
it was already known that the NLM ferrimagnetism is realized 
in the ground state of the kagome strip lattice 
in Fig.~\ref{fig1}(c). 
In the present study, 
we show that both the lattice in Fig.~\ref{fig1}(c) 
and the lattice in Fig.~\ref{fig1}(b) 
reveal the NLM ferrimagnetism in the ground state. 
Note also that the lattice shape at the edge 
under the open boundary condition 
depicted in Fig.~\ref{fig1}(c) is different 
from that in ref. \ref{kgm-stripe} 
[see Fig.~\ref{fig1}(b) in ref.~\ref{kgm-stripe}]. 
Thus, one can recognize that the results of the strip lattice 
with small width are irrespective of boundary conditions. 
We also present clearly the existence 
of the incommensurate modulation with long-distance periodicity 
in the local magnetizations of both models 
in Figs.~\ref{fig1}(b) and \ref{fig1}(c). 
Our numerical calculations suggest that 
the intermediate state of the 2D lattice 
in Fig. \ref{fig1}(a) is the NLM ferrimagnetism. 

This paper is organized as follows. 
In \S 2, we first present our numerical calculation methods.
In \S 3, we show the ground-state properties 
of the lattice depicted in Fig. \ref{fig1}(c) 
in finite-size clusters. 
In \S 4, we show the ground-state properties 
of the lattice depicted in Fig. \ref{fig1}(b). 
Sections 5 and 6 are devoted 
to discussion and summary, respectively. 

\section{Numerical Methods}
We employ two reliable numerical methods: 
the exact diagonalization (ED) method and 
the density matrix renormalization group (DMRG) 
method\cite{DMRG1, DMRG2}. 

The ED method is used to obtain precise physical quantities 
of finite-size clusters. 
This method does not suffer from the limitation 
of the cluster shape. 
It is applicable even to systems with frustration, 
in contrast to the quantum Monte Carlo (QMC) method 
coming across the so-called negative-sign problem for systems 
with frustration. 
The disadvantage of the ED method is the limitation 
that available system sizes are very small. 
Thus, we should pay careful attention 
to finite-size effects in quantities obtained by this method. 

On the other hand, 
the DMRG method is very powerful 
when a system is (quasi-)one-dimensional 
under the open boundary condition. 
The method can treat much larger systems 
than the ED method. 
Note that the applicability of the DMRG method is 
irrespective of whether or not the systems include frustrations. 
In the present research, 
we use the ``finite-system'' DMRG method. 

\section{Kagome Strip Lattice with Small Width}
In this section, we study the magnetic properties 
in the ground state of the $S=1/2$ Heisenberg model on 
the kagome strip lattice depicted in Fig. \ref{fig1}(c). 
The Hamiltonian of this model is given by
\begin{widetext}
\begin{eqnarray}
\label{Hamiltonian1}
\mathcal{H} &=&
  J_{1} \sum_{i}  [{\bf S}_{i,{\rm B}}\cdot {\bf S}_{i,{\rm C}}
                     + {\bf S}_{i,{\rm C}}\cdot {\bf S}_{i,{\rm A}^{\prime}}
                     + {\bf S}_{i,{\rm C}}\cdot {\bf S}_{i+1,{\rm A}}  
                     + {\bf S}_{i,{\rm C}}\cdot {\bf S}_{i+1,{\rm B}}]
\nonumber \\
&+&J_{2} \sum_{i}  [{\bf S}_{i,{\rm A}}\cdot {\bf S}_{i,{\rm B}}
                         +{\bf S}_{i,{\rm B}}\cdot {\bf S}_{i,{\rm A}^{\prime}}] 
+                     
J_{\rm F} \sum_{i} [{\bf S}_{i,{\rm A}}\cdot {\bf S}_{i+1,{\rm A}}
                     +{\bf S}_{i,{\rm A}^{\prime}}\cdot {\bf S}_{i+1, {\rm A}^{\prime}}],
\end{eqnarray}
\end{widetext}
where ${\bf S}_{i, \xi}$ is the $S=1/2$ spin operator 
at the $\xi$-sublattice site in the $i$-th unit cell. 
The positions of the four sublattices in a unit cell 
are denoted by A, ${\rm A}^{\prime}$, B, and C 
in Fig. \ref{fig1}(c).  
Energies are measured in units of $J_{1}$; 
we fixed $J_{1}=1$ hereafter. 
In what follows, we examine 
the region of $0<J_{2}/J_{1}< \infty$ 
in the case of $J_{\rm F}=-1$. 
Note that the number of total spin sites is denoted by $N$; 
thus, the number of unit cells is $N/4$. 

\begin{figure}[t]
\begin{center}
\includegraphics[width=6cm]{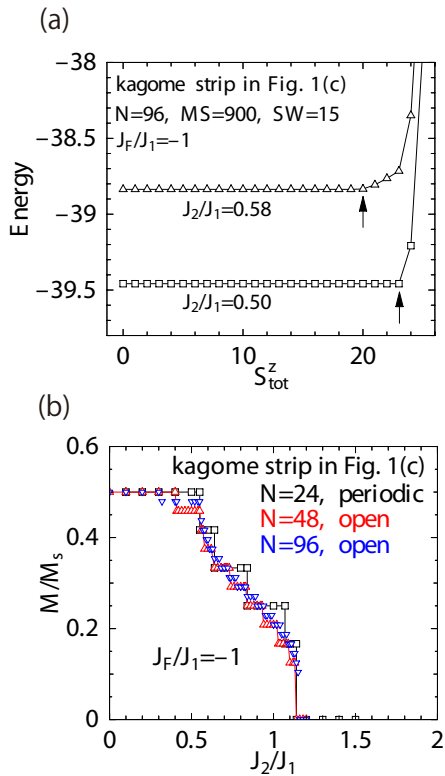}
\caption{(Color)(a) Dependence of the lowest energy 
on $S_{\rm tot}^{z}$. 
The results of $J_{2}/J_{1}=0.5$ and $0.58$ for the system size 
of $N=96$ are presented.
Arrows indicate the values of the spontaneous magnetization 
$M$ in each $J_{2}/J_{1}$. 
The position of an arrow is given 
by the highest $S_{\rm tot}^{z}$ value 
among those that give the common lowest energy. 
(b) $J_{2}/J_{1}$ dependence of $M/M_{\rm s}$ 
obtained from ED calculations 
for $N=24$ (black square) under the periodic boundary condition 
and DMRG calculation for $N=48$ (red triangle) and $96$ (blue inverted triangle) 
under the open boundary condition.
Note that for $N=96$ in (a) and (b), we use $MS=900$ and $SW=15$ 
and, that  
for $N=48$ in (b), we use $MS=400$ and $SW=15$. 
 }
\label{fig2}
\end{center}
\end{figure}

\begin{figure}[t]
\begin{center}
\includegraphics[width=6cm]{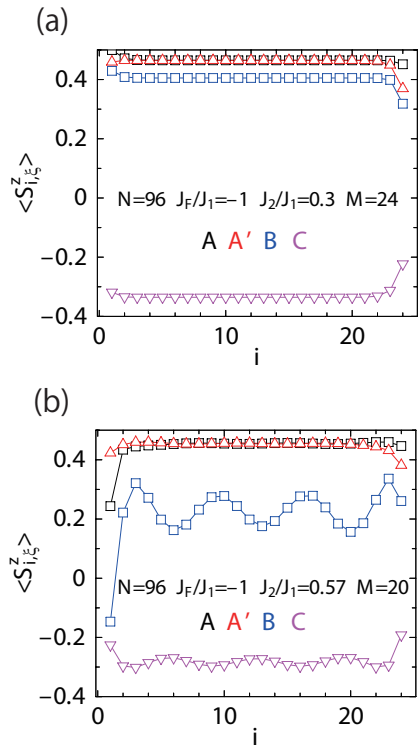}
\caption{(Color) Local magnetization $\langle S_{i,\xi}^{z} \rangle$ 
at each sublattice $\xi$; 
A (black square), ${\rm A}^{\prime}$ (red triangle), B (blue square), and 
C (purple inverted triangle).
Panels (a) and (b) show results 
for $J_{2}/J_{1}=0.3$ and $0.57$, respectively. 
These results are obtained from our DMRG calculations 
for $N=96$ ($i=1, 2, \cdots, 24$).
}
\label{fig3}
\end{center}
\end{figure}

We examine the $J_{2}/J_{1}$ dependence of the ratio 
$M/M_{\rm s}$, where $M$ and $M_{\rm s}$ are 
the spontaneous and 
saturation magnetizations, respectively. 
Let us explain the method used to determine 
$M$ as a function of $N$ and $J_{2}/J_{1}$. 
First, we calculate the lowest energy 
$E(J_{2}/J_{1}$, $S_{\rm tot}^{z}$, $N)$, 
where $S_{\rm tot}^{z}$ value is the $z$-component of the total spin. 
For example, 
the energies for various $S_{\rm tot}^{z}$ 
in the two cases of $J_{2}/J_{1}$
are presented in Fig. \ref{fig2}(a); 
the results are obtained by our DMRG calculations of the system 
of $N=96$ with the maximum number of retained states ($MS$) of 600 
and the number of sweeps ($SW$) of 15. 
The spontaneous magnetization $M(J_{2}/J_{1}$, $N$) is 
determined as the highest $S_{\rm tot}^{z}$ 
among those at the lowest common energy 
[see arrows in Fig. \ref{fig2}(a)]. 
Our results of the $J_{2}/J_{1}$ dependence of $M/M_{\rm s}$ 
are depicted in Fig. \ref{fig2}(b).
We find the existence of the intermediate magnetic phase 
of $0<M/M_{\rm s}<1/2$ between the LM ferrimagnetic phase 
of $M/M_{\rm s}=1/2$ and the nonmagnetic phase. 
In order to determine the spin state of this intermediate phase, 
we calculate the local magnetization 
$\langle S_{i, \xi}^{z} \rangle$, 
where $\langle A \rangle$ denotes the expectation value of
the physical quantity $A$ and 
$S_{i, \xi}^{z}$ is the $z$-component of ${\bf S}_{i, \xi}$. 
Figure \ref{fig3} depicts our results 
for a system size $N=96$ on the lattice depicted 
in Fig. \ref{fig1}(c) under the open boundary condition; 
Figs. \ref{fig3}(a) and \ref{fig3}(b) correspond to the case 
of the LM ferrimagnetic phase and that of the intermediate phase, 
respectively. 

The results clearly indicate  the existence 
of the incommensurate modulation with long-distance periodicity 
in the intermediate phase. 
We also confirm that the periodicities of the local magnetizations in the NLM ferrimagnetism in the present model 
depend on the $J_{2}/J_{1}$ value
but not on the length of the system, as reported 
in the case of ref. \ref{PF2}.

It is worth emphasizing the point that 
the intermediate phase commonly exists 
irrespective of the shape at the edge of the strip lattice 
when one compares the result of the present lattice depicted 
in Fig. \ref{fig1}(c) and that in ref. \ref{kgm-stripe}. 
Therefore, we can conclude that 
the intermediate phase exists and 
that it is NLM ferrimagnetic.

Although there exists an intermediate NLM ferrimagnetic phase 
in the case of the kagome strip lattice depicted 
in Fig. \ref{fig1}(c), 
we should note that there is a large discrepancy 
in dimensionality 
between the kagome strip lattice depicted in Fig. \ref{fig1}(c) 
and the 2D kagome lattice depicted in Fig. \ref{fig1}(a). 
In the next section, we treat the kagome strip lattice 
depicted in Fig. \ref{fig1}(b) whose width is larger than 
that of the kagome strip lattice depicted in Fig. \ref{fig1}(c).

\section{Kagome Strip Lattice with Large Width}
\subsection{Hamiltonian}
In this section, we study the ground-state properties 
of the $S=1/2$ Heisenberg model on the kagome strip lattice 
depicted in Fig. \ref{fig1}(b). 
The Hamiltonian of this model is given by
\begin{widetext}
\begin{eqnarray}
\label{Hamiltonian2}
\mathcal{H} &=&
  J_{1} \sum_{i}  [ {\bf S}_{i,{\rm B}}\cdot {\bf S}_{i,{\rm C}} 
                     + {\bf S}_{i,{\rm C}}\cdot {\bf S}_{i,{\rm D}}
                     + {\bf S}_{i,{\rm C}}\cdot {\bf S}_{i+1,{\rm A}}  
                     + {\bf S}_{i,{\rm C}}\cdot {\bf S}_{i+1,{\rm B}}
 \nonumber \\
                   &+& {\bf S}_{i,{\rm C}^{\prime}}\cdot {\bf S}_{i,{\rm B}^{\prime}}
                     + {\bf S}_{i,{\rm C}^{\prime}}\cdot {\bf S}_{i,{\rm A}^{\prime}}
                     + {\bf S}_{i,{\rm C}^{\prime}}\cdot {\bf S}_{i+1,{\rm D}}
                     +{\bf S}_{i,{\rm C}^{\prime}}\cdot {\bf S}_{i+1,{\rm B}^{\prime}} ]
\nonumber \\
&+&J_{2} \sum_{i}  [{\bf S}_{i,{\rm A}}\cdot {\bf S}_{i,{\rm B}}
                         +{\bf S}_{i,{\rm B}}\cdot {\bf S}_{i,{\rm D}}
                         +{\bf S}_{i,{\rm D}}\cdot {\bf S}_{i,{\rm B}^{\prime}}
                         +{\bf S}_{i,{\rm B}^{\prime}}\cdot {\bf S}_{i,{\rm A}^{\prime}}]                          
\nonumber \\
&+&                     
J_{\rm F} \sum_{i} [{\bf S}_{i,{\rm A}}\cdot {\bf S}_{i+1,{\rm A}}
                     +{\bf S}_{i,{\rm A}^{\prime}}\cdot {\bf S}_{i+1, {\rm A}^{\prime}}]
\nonumber \\
&-& h \sum_{i} [S^{z}_{i,{\rm A}}+S^{z}_{i,{\rm A}^{\prime}}+S^{z}_{i,{\rm B}}+S^{z}_{i,{\rm B}^{\prime}}+S^{z}_{i,{\rm C}}+S^{z}_{i,{\rm C}^{\prime}}+S^{z}_{i,{\rm D}}].
\end{eqnarray}
\end{widetext}
Here, the positions of seven sublattices are denoted 
by A, ${\rm A}^{\prime}$, B, ${\rm B}^{\prime}$, C, 
${\rm C}^{\prime}$, and D in Fig. \ref{fig1}(b). 
Note that the last term of eq. \ref{Hamiltonian2} is the Zeeman term.
The number of spin sites is denoted by $N$. 
The number of unit cells is $N/7$; 
we consider $N/14$ as an integer. 
We mainly use the DMRG method 
for investigating the magnetic properties 
in the ground state of this Q1D system 
under the open boundary condition. 
We also investigate the properties 
under the periodic boundary condition by the ED method, 
although the size treated by this method is only 
in the case of $N=28$.  
Hereafter, we consider $J_{1}=1$ as an energy scale and 
we investigate the region of $0<J_{2}/J_{1}<\infty$ 
in the case of $J_{\rm F}=-1$. 

\subsection{Phase diagram}
\begin{figure}[h]
\begin{center}
\includegraphics[width=5.6cm]{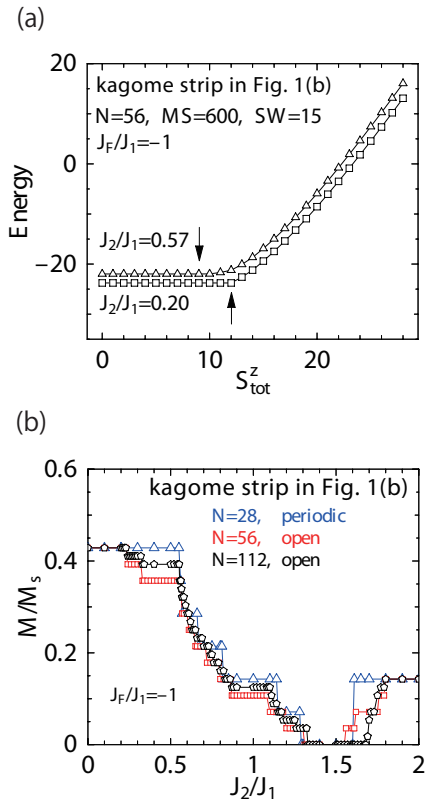}
\caption{(Color)(a) Lowest energy in each subspace divided by $S_{\rm tot}^{z}$.
The results of the DMRG calculations obtained when the system size is $N=56$ for $J_{2}/J_{1}=0.20$ and 0.57 are presented. 
Arrows indicate the spontaneous magnetization $M$ for a given $J_{2}/J_{1}$.
(b) $J_{2}/J_{1}$ dependence of $M/M_{\rm s}$ obtained from the ED calculations for $N=$28 (blue triangle) 
under the periodic boundary condition and the DMRG calculations for $N=$56 (red square) and 112 (black pentagon) under the open boundary condition. 
Note that for $N=$ 56 in (a) and (b), we use $MS=$ 600 and $SW=$15, and that for $N=112$ in (b), we use $MS\geq$ 900 and $SW=$15.
Here, $MS$ denotes the maximum number of retained states and $SW$ the number of sweeps used in DMRG calculations.
}
\label{fig4}
\end{center}
\end{figure}

\begin{figure}[h]
\begin{center}
\includegraphics[width=5.6cm]{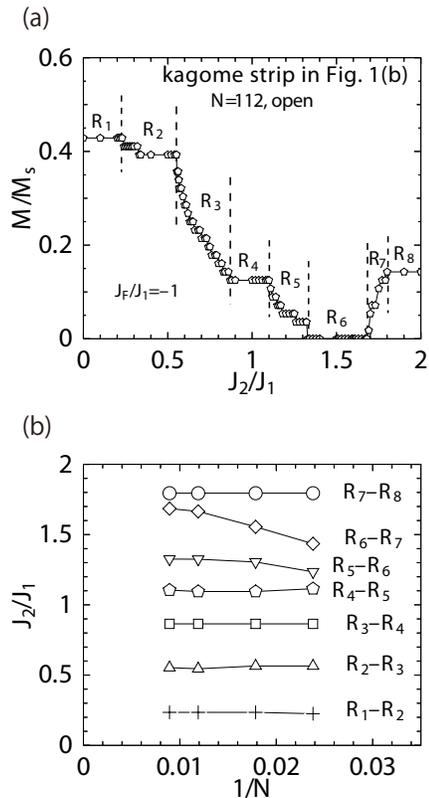}
\caption{(a) Definitions of the ${\rm R}_{1}$-${\rm R}_{8}$ regions in the case of $N=112$ under the open boundary condition:
${\rm R}_{1}$ is the region of $M/M_{\rm s}=3/7$, 
${\rm R}_{2}$ is the region of $11/28\leq M/M_{\rm s} <3/7$,
${\rm R}_{3}$ is the region of $1/8 < M/M_{\rm s} <11/28$, 
${\rm R}_{4}$ is the region of $M/M_{\rm s} =1/8$,
${\rm R}_{5}$ is the region of $0< M/M_{\rm s} <1/8$,
${\rm R}_{6}$ is the region of $M/M_{\rm s} =0$,
${\rm R}_{7}$ is the region of $0 < M/M_{\rm s} <1/7$, and 
${\rm R}_{8}$ is the region of $M/M_{\rm s} =1/7$.
Dashed lines indicate the boundaries of these regions.
(b) Size dependences of boundaries under the open boundary condition.
The results presented are those of $N=$42, 56, 84, and 112 from DMRG calculations.
The curve line named ${\rm R}_{l}$-${\rm R}_{l+1}$ indicates the boundary line between the ${\rm R}_{l}$ and ${\rm R}_{l+1}$ regions, where $l$ is integer.
}
\label{fig5}
\end{center}
\end{figure}

First, let us examine the $J_{2}/J_{1}$ dependence 
of the ratio $M/M_{\rm s}$ in the absence 
of the external magnetic field $h$.  
The procedure for determining $M$ is 
the same as that mentioned in \S 3 [see also Fig. \ref{fig4}(a)]. 
We present our results of the spontaneous magnetization 
in Fig. \ref{fig4}(b). 
We successfully observe the intermediate-magnetization region 
irrespective of the boundary conditions. 
A careful observation of Fig. \ref{fig4}(b) enables us 
to observe the eight regions at least 
in the finite-size system. 
As a matter of convenience, hereafter, 
we call these regions ${\rm R}_{1}$, ${\rm R}_{2}$, $\cdots$, 
${\rm R}_{7}$, and ${\rm R}_{8}$. 
In the case of $N=112$ under the open boundary condition,
for example, Fig. \ref{fig5}(a) illustrates 
the regions 
${\rm R}_{1}$ to ${\rm R}_{8}$: 
${\rm R}_{1}$ is the region of $M/M_{\rm s}=3/7$, 
${\rm R}_{2}$ is the region of $11/28\leq M/M_{\rm s} <3/7$,
${\rm R}_{3}$ is the region of $1/8 < M/M_{\rm s} <11/28$, 
${\rm R}_{4}$ is the region of $M/M_{\rm s} =1/8$,
${\rm R}_{5}$ is the region of $0< M/M_{\rm s} <1/8$,
${\rm R}_{6}$ is the region of $M/M_{\rm s} =0$,
${\rm R}_{7}$ is the region of $0 < M/M_{\rm s} <1/7$,
and ${\rm R}_{8}$ is the region of $M/M_{\rm s} =1/7$.
Here, the dashed lines in Fig. \ref{fig5}(a) indicate 
the boundaries of these regions. 

It should be noted that 
the values of $M/M_{\rm s}$ in the ${\rm R}_{4}$ region 
and that at the lower edge of the ${\rm R}_{2}$ region 
change with increasing $N$, as shown in Fig. \ref{fig4}(b); 
the former value is $M/M_{\rm s}=(N-14)/7N$ and 
the latter value is $M/M_{\rm s}=(3N-28)/7N$. 
These changes due to the increase in system size 
come from the finite-size effect. 
We find that the value of $M/M_{\rm s}$ 
in the ${\rm R}_{4}$ region 
under the open boundary condition increases and 
approaches the value of $M/M_{\rm s}=1/7$ 
when $N$ increases.
Furthermore, 
the magnetization value in the ${\rm R}_{4}$ region is 
$M/M_{\rm s}=1/7$ in the case of $N=28$ 
under the periodic boundary condition. 
Therefore, it is expected 
that the value of $M/M_{\rm s}$ in the ${\rm R}_{4}$ region 
is $1/7$ in the thermodynamic limit. 
We also confirm that 
the value of $M/M_{\rm s}$ at the lower edge 
of the ${\rm R}_{2}$ region gradually increases and 
approaches the value of $M/M_{\rm s}=3/7$ 
with increasing $N$.
In addition, we cannot confirm the ${\rm R}_{2}$ region 
in the case of $N=28$ under the periodic boundary condition. 
These circumstances indicate a possibility that 
the ${\rm R}_{2}$ region merges 
with the ${\rm R}_{1}$ region of $M/M_{\rm s}=3/7$ 
in the thermodynamic limit.

\begin{figure}[h]
\begin{center}
\includegraphics[width=6cm]{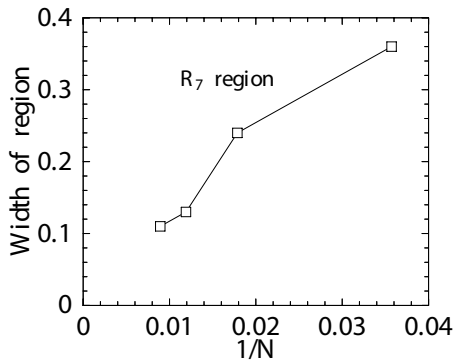}
\caption{Size dependence of the width of the ${\rm R}_{7}$ region.
It is difficult to determine whether the ${\rm R}_{7}$ region survives in the thermodynamic limit.
}
\label{fig6}
\end{center}
\end{figure}

Next, to determine whether 
each region survives in the thermodynamic limit, 
we study the size dependences of the boundaries 
between the regions under the open boundary condition. 
Figure \ref{fig5}(b) shows the results 
of $N=$42, 56, 84, and 112 from DMRG calculations. 
Note here that we define ${\rm R}_{2}$ as 
the region of $(3N-28)/7N<M/M_{\rm s}<3/7$ and 
${\rm R}_{4}$ as the region 
of $M/M_{\rm s}=(N-14)/7N$ in the finite-size system. 
One can find immediately from Fig. \ref{fig5}(b) that 
all regions, except the ${\rm R}_{7}$ region, survive 
in the limit $N \rightarrow \infty$. 
To determine whether the ${\rm R}_{7}$ region survives 
in the thermodynamic limit, we investigate the size dependence 
of the width of the ${\rm R}_{7}$ region in Fig. \ref{fig6}. 
This plot shows us that the width of the ${\rm R}_{7}$ region 
decreases with increasing $N$. 
It is difficult to determine whether 
the ${\rm R}_{7}$ region survives in the thermodynamic limit. 
The convex downward behavior is observed for large sizes 
so that the region might survive; however, 
the observed behavior may be one of the serious finite-size effects. 
The issue of establishing the presence or absence 
of the ${\rm R}_{7}$ region should be clarified 
in future studies.

\subsection{Magnetic properties in each region}
\begin{figure}[h]
\begin{center}
\includegraphics[width=6cm]{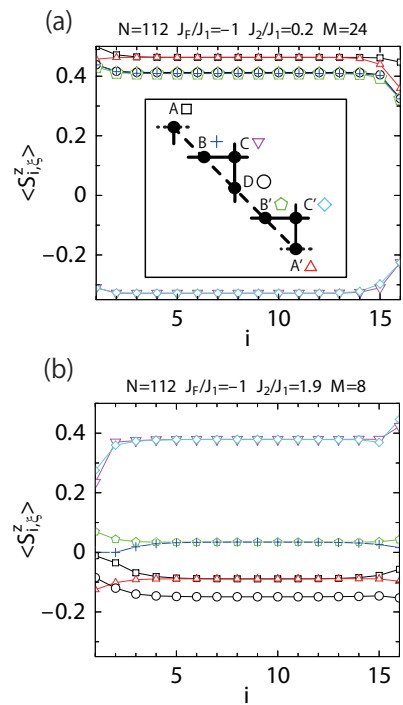}
\caption{(Color) Local magnetization  $\langle S_{i,\xi }^{z} \rangle$ at each sublattice $\xi$.
Panels (a) and (b) show results for $J_{2}/J_{1}=$0.2 and 1.9, respectively.
These results are obtained from our DMRG calculations for $N=112$ ($i=$1, 2, $\cdots$, 16).
The inset in the panel (a) presents the correspondence relationship between each colored symbol and each sublattice $\xi$ used in Figs. \ref{fig7}, \ref{fig9}, and \ref{fig10}.
} 
\label{fig7}
\end{center}
\end{figure}

In this subsection, we investigate the local magnetization 
$\langle S_{i,\xi }^{z} \rangle$ to study 
the magnetic structures in various regions 
except the ${\rm R}_{2}$ and ${\rm R}_{6}$ regions.
Note that we calculate $\langle S_{i,\xi }^{z} \rangle$ 
within the subspace of the highest $S_{\rm tot}^{z}$ 
corresponding to the spontaneous magnetization 
$M$ when $J_{2}/J_{1}$ is given. 
Considering the fact that the present lattice depicted 
in Fig.~\ref{fig1}(b) has seven sublattices in the system, 
we will use different colors or symbols for each sublattice $\xi$ 
for presenting our results of $\langle S_{i,\xi }^{z} \rangle$, 
as depicted in the inset of Fig. \ref{fig7}(a); 
we use a black square for $\xi=$A, 
a red triangle for $\xi={\rm A}^{\prime}$, 
a blue cross for $\xi={\rm B}$, 
a green pentagon for $\xi={\rm B}^{\prime}$, 
a purple inverted triangle for $\xi=$C, 
an aqua diamond for $\xi={\rm C}^{\prime}$, 
and a black circle for $\xi=$D. 

First, we examine the ${\rm R}_{1}$ and ${\rm R}_{8}$ regions. 
We present our DMRG results of $\langle S_{i,\xi }^{z} \rangle$ 
of the system of $N=112$ in Figs. \ref{fig7}(a) 
and \ref{fig7}(b) for $J_{2}/J_{1}=0.2$ and 1.9, respectively. 
In Fig. \ref{fig7}(a), 
we observe the uniform behavior of upward-direction spins 
at the sublattice sites A, ${\rm A}^{\prime}$, B, ${\rm B}^{\prime}$, 
and D, and downward-direction spins at the sublattice sites C and 
${\rm C}^{\prime}$. 
In Fig. \ref{fig7}(b), we also observe the uniform behavior 
of upward-direction spins at the sublattice sites B, 
${\rm B}^{\prime}$, C, and ${\rm C}^{\prime}$, and 
downward-direction spins at the sublattice sites A, 
${\rm A}^{\prime}$, and D. 
Therefore, we conclude that the LM ferrimagnetic states 
are realized 
in the regions of ${\rm R}_{1}$ and ${\rm R}_{8}$.

\begin{figure}[h]
\begin{center}
\includegraphics[width=6cm]{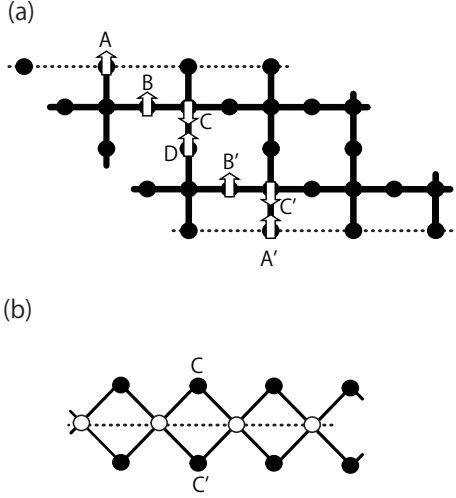}
\caption{(a) Kagome strip lattice depicted in Fig. \ref{fig1}(b) in the limit of $J_{2}/J_{1}=0$. 
White arrowheads denote the classical spin configuration in the LM ferrimagnetic state of $M/M_{\rm s}=3/7$.
(b) Kagome strip lattice depicted in Fig. \ref{fig1}(b) in the limit of $J_{2} \rightarrow \infty$.
White circles represent the effective $S=1/2$ spins formed by five $S=1/2$ spins 
at the sublattices $\xi=$A, ${\rm A}^{\prime}$, B, ${\rm B}^{\prime}$, and D. 
The black bold and black dotted lines show the antiferromagnetic and ferromagnetic interactions, respectively.
}
\label{fig8}
\end{center}
\end{figure}

Our understanding of the origins of these LM ferrimagnetic phases 
is based on the Marshall-Lieb-Mattis (MLM) theorem. 
In the case of $J_{2}/J_{1}=0$, no frustration occurs; 
thus, the spin state depicted in Fig. \ref{fig8}(a) 
is realized. 
This state shows the LM ferrimagnetism with $M/M_{\rm s}=3/7$.  
The ${\rm R}_{1}$ region is directly connected 
to the LM ferrimagnetic state of $J_{2}/J_{1}=0$. 
Therefore, the ${\rm R}_{1}$ region of $M/M_{\rm s}=3/7$ is 
regarded as the LM ferrimagnetic phase. 
In the limit $J_{2} \rightarrow \infty$, on the other hand, 
the present model becomes equal to a model 
of an $S=1/2$ diamond chain depicted in Fig. \ref{fig8}(b). 
The value of magnetization takes $M/M_{\rm s}=1/7$ 
in the ground state of this diamond chain 
according to the MLM theorem. 
Therefore, the ${\rm R}_{8}$ region of $M/M_{\rm s}=1/7$ is 
regarded as the LM ferrimagnetic phase. 

\begin{figure}[h]
\begin{center}
\includegraphics[width=6.0cm]{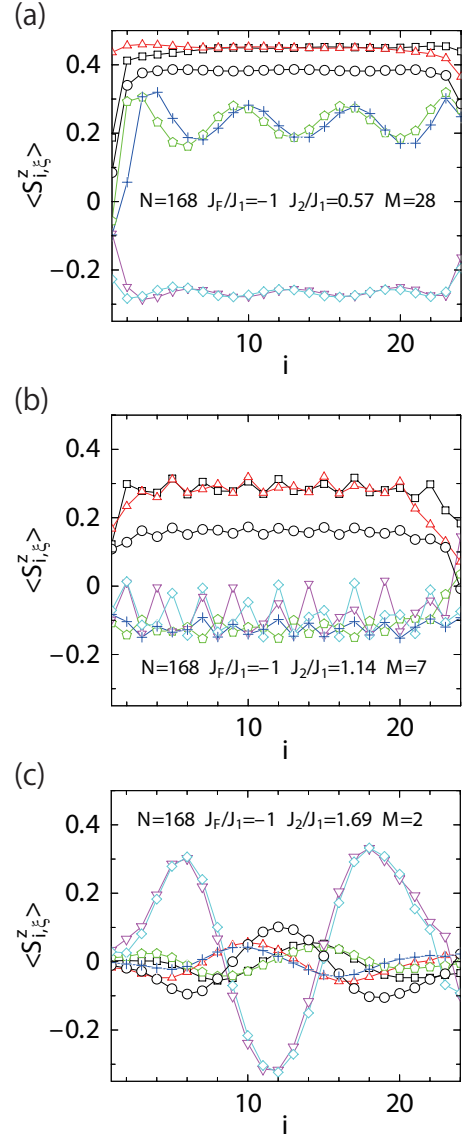}
\caption{(Color) Local magnetization  $\langle S_{i,\xi }^{z} \rangle$ at each sublattice $\xi$.
The correspondence relationship between each colored symbol and each sublattice $\xi$ is described in the inset in Fig. \ref{fig7}(a).
Panels (a), (b), and (c) show results for $J_{2}/J_{1}=$0.57, 1.14, and 1.69, respectively.
These results are obtained from our DMRG calculations for $N=168$ ($i=$1, 2, $\cdots$, 24).
}
\label{fig9}
\end{center}
\end{figure}

Next, we investigate the ${\rm R}_{3}$, ${\rm R}_{5}$, 
and ${\rm R}_{7}$ regions. 
Our results obtained from the DMRG calculations of $N=168$ are 
depicted in Figs. \ref{fig9}(a)-\ref{fig9}(c) 
for $J_{2}/J_{1}=0.57$, 1.14, and 1.69, respectively.
We clearly observe incommensurate modulations 
with long-distance periodicities 
in every case in Fig.~\ref{fig9}. 
In addition, we confirm from Fig. \ref{fig4}(b) that 
the ratio $M/M_{\rm s}$ changes gradually 
with the variation in $J_{2}/J_{1}$ 
in the R$_3$, R$_5$, and R$_7$ regions.  
Since the widths of the R$_3$ and R$_5$ regions 
survive in the thermodynamic limit 
as was clarified in the previous subsection, 
we conclude that 
the ${\rm R}_{3}$ and ${\rm R}_{5}$ regions 
are NLM ferrimagnetic phases. 
Although it is unclear whether the ${\rm R}_{7}$ region 
survives in the thermodynamic limit, 
this region is an NLM ferrimagnetic phase 
if it survives. 

\begin{figure}[h]
\begin{center}
\includegraphics[width=6cm]{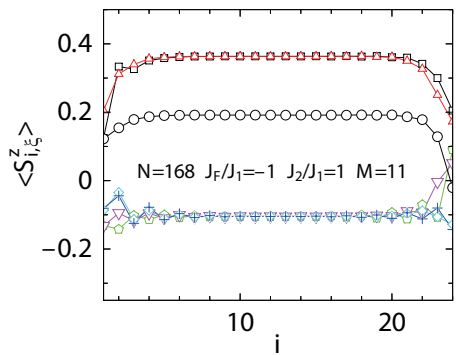}
\caption{(Color) Local magnetization  $\langle S_{i,\xi }^{z} \rangle$ at each sublattice $\xi$.
This result is obtained from our DMRG calculations for $J_{2}/J_{1}=1$ in the system of $N=168$ ($i=$1, 2, $\cdots$, 24).
The correspondence relationship between each colored symbol and each sublattice $\xi$ is described in the inset in Fig. \ref{fig7}(a).
}
\label{fig10}
\end{center}
\end{figure}

\begin{figure}[h]
\begin{center}
\includegraphics[width=6cm]{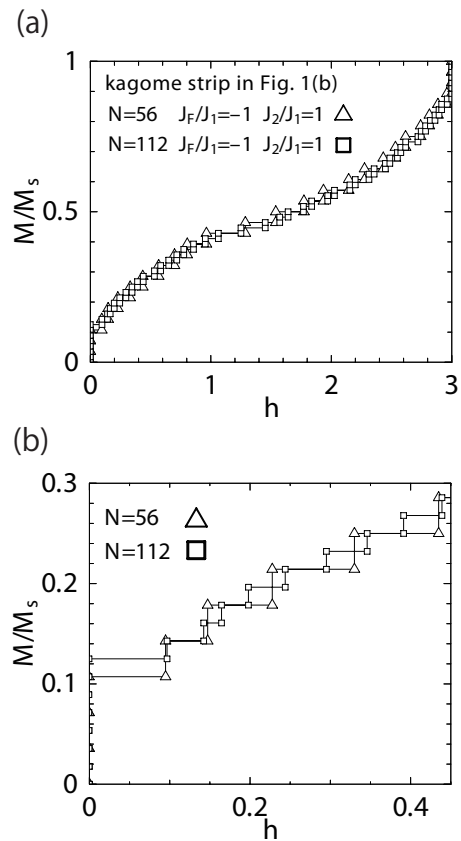}
\caption{Magnetization curve in the ${\rm R}_{4}$ region.
Panel (a) is obtained from the DMRG calculations of $J_{2}/J_{1}=1$ for $N=56$ (triangle) and $N=112$ (square).
Panel (b) is obtained by zooming the region near $h=0$ in panel (a).
}
\label{fig11}
\end{center}
\end{figure}

Finally, in this subsection, we examine the ${\rm R}_{4}$ region.
Our result of $\langle S_{i,\xi }^{z} \rangle$ 
for $J_{2}/J_{1}=1$ in the system of $N=168$ is 
depicted in Fig. \ref{fig10}. 
We do not detect the incommensurate modulation 
in this ${\rm R}_{4}$ region. 
In addition, we confirm from Fig. \ref{fig4}(b) that 
the ratio $M/M_{\rm s}$ in the ${\rm R}_{4}$ region 
does not change with the variation in $J_{2}/J_{1}$ 
in contrast to the cases in the R$_3$ and R$_5$ regions. 
These circumstances suggest that 
the ${\rm R}_{4}$ region is the LM ferrimagnetic phase. 
However, one cannot speculate the spin state 
from the result of $\langle S_{i,\xi }^{z} \rangle$ 
because it is difficult to clarify the spin state 
on the basis of the results of determining whether the spins 
are up or down, as was successfully observed in the ${\rm R}_{8}$ region. 
We further investigate 
the magnetization curve in this region 
to know whether the ${\rm R}_{4}$ region is 
the LM ferrimagnetic phase. 
The magnetization curves of $J_{2}/J_{1}=1$ 
for $N=56$ and $N=112$ calculated 
by the DMRG method are presented in Fig. \ref{fig11}(a)\cite{comment}.
Figure \ref{fig11}(b) is obtained by zooming 
the region of $h$ near $h=0$. 
One can observe the existence of the magnetization plateaus 
at the height of the spontaneous magnetization 
for both system sizes. 
We also confirm that the difference in the width 
of the plateau between $N=112$ and 56 is very small. 
These features indicate that 
the spin gap exists in the ${\rm R}_{4}$ region 
in the thermodynamic limit. 
If the ${\rm R}_{4}$ region is the NLM ferrimagnetic phase, 
no spin gap is present in this region. 
(It was reported in ref. \ref{Hida} that 
the NLM ferrimagnetism is gapless 
as a response to a uniform magnetic field.) 
Therefore, we conclude that 
the ${\rm R}_{4}$ region is the LM ferrimagnetic phase. 

\section{Discussion}
We discuss the relationships 
between the interval $0< J_{2}/J_{1}\le 1$ 
of the kagome strip lattice depicted in Fig. \ref{fig1}(b) and 
those of the spatially anisotropic kagome lattice 
studied in ref. \ref{kagome-2D}, 
while we consider the case of another new lattice, 
with a larger but finite width 
along the direction perpendicular to the bonds 
of interaction $J_{\rm F}$ in Fig. \ref{fig1}(b), 
namely, the strip width. 

The ratio $M/M_{\rm s}$ of the ${\rm R}_{1}$ 
and ${\rm R}_{2}$ regions is commonly $M/M_{\rm s}=3/7$ 
in the thermodynamic limit of the lattice depicted 
in Fig. \ref{fig1}(b). 
The difference of $M/M_{\rm s}=3/7$ from $M/M_{\rm s}=1/3$ 
in the case of the LM ferrimagnetic phase of 
the spatially anisotropic kagome lattice 
in ref. \ref{kagome-2D} is attributed to the finiteness of the strip width. 
Thus, the ratio $M/M_{\rm s}$ approaches $M/M_{\rm s}=1/3$ 
when the strip width increases; 
the ${\rm R}_{1}$ and ${\rm R}_{2}$ regions of the present model 
depicted in Fig. \ref{fig1}(b) correspond to the LM ferrimagnetic 
phase of the spatially anisotropic kagome lattice 
studied in ref. \ref{kagome-2D}.  

The ratio $M/M_{\rm s}=1/7$ at $J_{2}/J_{1}=1$ 
in the present model depicted in Fig. \ref{fig1}(b) 
may be related to the fact that 
the model includes seven sublattices, 
namely, the strip width is finite. 
At least at $J_{2}/J_{1}=1$, 
on the other hand, 
it is widely believed that 
the spontaneous magnetization disappears 
in the limit of infinite width\cite{Lecheminant,
Waldtmann,Hida_kagome,Cabra,Honecker0,Honecker1,Cepas,
Spin_gap_Sindzingre,HN_Sakai2010,Sakai_HN_PRBR,HN_Sakai2011}.   
Although the relationship should be clarified 
in the examination of the case of the lattice 
with an even larger strip width, 
there are two possibilities of the behavior 
near the case of $J_{2}/J_{1}=1$.  
One is the case when 
the ratio $M/M_{\rm s}$ decreases 
with increasing strip width and finally vanishes, 
while the LM ferrimagnetic phase, such as ${\rm R}_4$, 
survives in systems with larger strip widths. 
In this case, the ${\rm R}_{3}$ region corresponds 
to the NLM phase of the two-dimensional model 
on the spatially anisotropic kagome lattice. 
In the other case, 
the LM ferrimagnetic phase, such as ${\rm R}_4$, 
becomes narrower with increasing strip width, 
while the ratio $M/M_{\rm s}$ does not vanish; 
finally, the ${\rm R}_3$ and ${\rm R}_5$ regions 
merge with each other. 
In this case, the value of $J_2/J_1$ at the boundary 
between the ${\rm R}_5$ and ${\rm R}_6$ regions 
decreases across $J_2/J_1=1$.  
In any cases, it is important to note that from 
our finding of an intermediate phase 
in all the three cases in Fig.~\ref{fig1} 
between the ferrimagnetic phase and the nonmagnetic phase, 
the phase is considered to exist irrespective of the strip width. 

Finally, one should note that 
the intermediate phase between the Lieb-Mattis ferrimagnetic 
and nonmagnetic states is observed in other cases. 
However, this phase is not always ferrimagnetic. 
One of the cases observed is the case of the three-leg ladder system 
forming a strip lattice obtained by cutting off from 
the spatially anisotropic triangular lattice 
in ref.~\ref{Abe}, 
in which the properties of the intermediate phase were unclear. 
Sakai's unpublished study suggests that the intermediate phase 
is nematic\cite{Sakai_private_com}. 
Careful examinations are required to investigate 
such an intermediate phase if it is found. 

\section{Summary}
We have studied the ground-state properties 
of the $S=1/2$ Heisenberg models on the kagome strip lattices 
depicted in Figs. \ref{fig1}(b) and \ref{fig1}(c) 
by the ED and DMRG methods. 
As a common phenomenon in the ground state of both cases, 
we have confirmed the existence 
of the non-Lieb-Mattis ferrimagnetism 
between the Lieb-Mattis ferrimagnetic phase and 
the nonmagnetic phase. 
We have clearly found 
incommensurate modulations 
with long-distance periodicity 
in the non-Lieb-Mattis ferrimagnetic state. 
The occurrence of the non-Lieb-Mattis ferrimagnetism 
irrespective of strip width strongly suggests that 
the intermediate state found 
in the case of the spatially anisotropic kagome lattice 
in two dimensions is the non-Lieb-Mattis ferrimagnetism.

\section*{Acknowledgments}
We thank Prof. Toru Sakai for letting us know his 
unpublished results. 
This work was partly supported  
by Grants-in-Aid 
(Nos. 20340096, 23340109, 23540388, and 24540348) 
from the Ministry of Education, Culture, Sports, 
Science and Technology of Japan.
This work was partly supported by 
a Grant-in-Aid (No. 22014012) 
for Scientific Research and Priority Areas 
``Novel States of Matter Induced by Frustration'' 
from the Ministry of Education, Culture, Sports, Science 
and Technology of Japan. 
Diagonalization calculations in the present work were 
carried out using TITPACK Version 2 coded by H. Nishimori.
DMRG calculations were carried out 
using the ALPS DMRG application\cite{ALPS}.
Some computations were performed 
using the facilities of 
the Supercomputer Center, Institute for Solid 
State Physics, University of Tokyo.


\begin{thebibliography}{99}
\bibitem{comment}
One can find the macroscopic jump in this magnetization curve near the saturation field as well as in the case of some frustrated systems\cite{kagome-strip3, localized-magnon}. 
It is known that the occurrence mechanism of the magnetization jump can be understood from the viewpoint of the localized-magnon picture (see refs. \ref{magnon1}, \ref{magnon2}, and \ref{magnon3}).
\end{thebibliography}
\end{document}